\documentclass[aps,prl, twocolumn, groupedaddress]{revtex4-1}

\usepackage{graphicx}
\usepackage{float,amsmath,verbatim,amssymb,bm}

\begin{document}


\title{Full-scale ab initio 3D PIC simulations of  an all-optical radiation reaction configuration at $10^{21}\mathrm{W/cm^2}$}
\preprint{twocolumn}

\author{Marija Vranic, Joana L. Martins, Jorge Vieira, Ricardo A. Fonseca and Luis O. Silva}
\affiliation{}


\date{\today}

\begin{abstract}
Using full-scale 3D particle-in-cell simulations we show that the radiation reaction dominated regime can be reached in an all optical configuration through the collision of a $\sim$1 GeV laser wakefield accelerated (LWFA) electron bunch with a counter propagating laser pulse. In this configuration radiation reaction significantly reduces the energy of the particle bunch, thus providing clear experimental signatures for the process with currently available lasers. We also show that the transition between classical and quantum radiation reaction could be investigated in the same configuration with laser intensities of $10^{24}\mathrm{W/cm^2}$.
\end{abstract}
\pacs{}

\maketitle


Radiation reaction is the change of momentum of a charged particle while it radiates. This apparently simple problem has many subtleties and it remains a  
long-standing fundamental question yet to be fully understood. The Lorentz-Abraham-Dirac (LAD) equation was an attempt to self-consistently account for radiation reaction in the classical regime \cite{LAD}. However, this equation contains unphysical runaway solutions and violates the causality principle, which gave rise to various alternative models to account for this effect \cite{LLbook, model1bell, model3sokolov},  and in particular the model proposed by Landau and Lifshitz (LL)  \cite{LLbook}. It contains all the physical solutions of the LAD equation \cite{spohn}, is free of the problems aforementioned \cite{rorlich_1,spohn}, and is therefore a strong candidate to describe the classical radiation reaction.  There is also a strong debate about the threshold at which the quantum effects prevail \cite{Ruhl_threshold, model1bell, model3sokolov}. Moreover, the extremely high laser intensities required to enter the radiation reaction dominated regime have hindered the experimental clarification of these issues, synchrotron radiation remaining the only confirmation up to date. 

It is then clear that the experimental demonstration of the radiation reaction regime and its signatures is of paramount importance. 
Schemes to detect radiation reaction have recently been proposed  \cite{capturePiazza, ThomasRR}, but a configuration where such regime can be accessed remains to be identified and tested. In this paper, we identify a radiation reaction dominated regime easily achievable with current state-of-the-art lasers ($10^{21}\mathrm{W/cm^2}$). Using 3-dimensional full-scale \emph{ab initio} particle-in-cell simulations, we explore an all-optical scheme based on head-on scattering a laser pulse off a LWFA electron bunch. Electron bunches with 1.5 GeV energy and 100 pC charge have already been obtained experimentally in 1 cm long plasmas  \cite{Lwfa_gev_1, Lwfa_gev_2, Lwfa_gev_3}  in LWFA.  Also, it has recently been confirmed experimentally that LWFA electrons can produce few hundred keV radiation in head-on interaction with laser pulses\cite{MalkaNature}, for a setup in the regime where the radiation reaction was still negligible. This clearly indicates that a head-on laser-LWFA electron beam configuration is feasible.  
It is the purpose of this work to explore the classical radiation reaction effects for currently available laser and plasma parameters, and to identify signatures of this process. We find that the head-on collision between a LWFA generated electron bunch with energy 0.5 - 1.5 GeV, and a counter-propagating scattering laser pulse of intensity $10^{20}-10^{22}\mathrm{W/cm^2}$ (Fig. 1) leads to significant electron beam energy loss and energy-spread reduction that can be easily detected in an experiment. The interaction is accompanied by hard X-ray emission. This configuration can produce more than $10^{11}$ photons, with energies in 10 -100 keV range.

We start by analytically estimating how much energy the electrons loose during the interaction with the scattering laser pulse. For the sake of completeness, we examine the total radiated power (averaged over the solid angle) for a single electron undergoing Compton scattering in a plane electromagnetic wave \cite{RPastro} given by: 
\begin{equation}\label{radP}
P=-\frac{d(\gamma mc^2)}{dt}=c\sigma_C\gamma^2(1-\beta \cos \theta)^2 U_{PH}
\end{equation}
where $\sigma_C$ is the Compton cross section (for the case of ultra relativistic electrons, the Compton scattering cross section converges to the Thompson cross section $\sigma_C\approx \sigma_T=8\pi r_0^2/3$), $r_0=e^2/mc^2$ is the classical electron radius, $e$ is the elementary charge, $m$ is the electron mass, $\gamma$ is the electron Lorentz factor, $\theta$ is the angle between
the $\mathbf{k}$ vector of the counter-propagating electromagnetic wave and $\bm{\beta}$, the electron velocity normalized to $c$, and $U_{PH}=(E^2+B^2)/8\pi$ is the energy density of the electromagnetic field. Equation \eqref{radP} is valid for $\gamma \hbar \omega_0<<mc^2$ ($ \omega_0$ is the frequency of the electromagnetic wave), i.e. when in its rest frame the electron
still undergoes the classical Thomson scattering.
For an ultrarelativistic electron $|\bm{\beta}|\approx1$, and Eq. \eqref{radP} becomes 

\begin{equation}\label{energyloss}
\frac{d\gamma}{dt}=-\frac{e^2\omega_0^2}{3mc^3}(1-\cos\theta)^2a_0^2 \gamma^2
\end{equation}
where  $a_0=eA/mc^2$ is the normalized vector potential. In a plane wave with constant amplitude, Eq. \eqref{energyloss} can be integrated to give
$\gamma(t)=\gamma_0/(1+\alpha t \gamma_0)$, where $ \alpha=(e^2\omega_0^2/3mc^3)(1-\cos\theta)^2a_0^2.$
Assuming the laser pulse is a plane wave with a temporal envelope $a_0(t)$, integration of Eq. \eqref{energyloss} yields an estimate for the final electron energy after interacting with the scattering laser: 

\begin{equation}\label{efinal}
\gamma_f=\frac{\gamma_0}{1+k \gamma_0}, \quad k=(1-\cos\theta)^2\frac{\eta}{3}\frac{e^2\omega_0^2}{mc^3}a_0^2 \tau_0
\end{equation}

where $\gamma_0$ and $\gamma_f$ are the initial and the final relativistic factor of the electron, $\tau_0$ is the scattering laser pulse duration at FWHM in the laser fields, and $a_0$ is the peak normalized vector potential of the scattering laser and where the crossing time is $\approx \tau_0/2$. The factor $\eta\approx0.4$ accounts for the different temporal profiles where, for instance, $\eta=0.375$ for an envelope $a_0(t)=a_0 \sin^2(t\pi/2\tau_0)$, and $\eta=0.392$ for the polynomial envelope we have used in the simulations (described later along with other simulation parameters). The coefficient $k$ depends only on the scattering laser parameters, and can be written in a more convenient way as
$k=1.22\times10^{-4} I_0\left[10^{22} \mathrm{W}/\mathrm{cm^2}\right]\tau_0[\mathrm{fs}](1-\cos \theta)^2$, where $I_0$ is the scattering laser peak intensity. We can now relate the properties of the electron bunch with those of the plasma when the laser drives the LWFA  \cite{LWFAscailing}. The estimated output electron energy from a LWFA is determined by $\gamma_0=(2/3) \left(\omega_{LD}/\omega_p \right)^2 a_{LD}$, where $\omega_{LD}$ and $ a_{LD}$ are the frequency and the normalized vector potential of the laser driver, and $\omega_p=(4\pi n_ee^2/m)^{1/2}$ is the electron plasma frequency ($n_e$ stands for electron density). For a head-on collision $1-\cos\theta\approx2$, and Eq. \eqref{efinal} shows that an electron beam looses 50\% of the energy in the interaction with the scattering laser when $k\gamma_0=6.6  \times 10^{-2} I_0\left[\mathrm{10^{22}W/cm^2} \right]  \ \tau_0[10 \mathrm{fs}] \times \sqrt{I_{LD}\left[\mathrm{10^{22}W/cm^2} \right] }  \  \left(\omega_{LD}/\omega_p \right)^2\simeq 1$
which for typical LWFA parameters $\omega_{LD}\gtrsim 10-100\  \omega_p$ can be easily achieved (here $I_{LD}$ is the intensity of the driver laser).

Equation \eqref{efinal} also shows that  the electron energy spectrum becomes significantly narrower during Compton scattering in this configuration since faster electrons radiate a larger percentage of their energy than the less energetic electrons. For a quasi-monoenergetic electron beam the relative energy spread decreases at the same rate as the mean energy viz.
\begin{equation}\label{enspread}
\frac{\delta\gamma_f}{\gamma_f}=\frac{1}{1+k\gamma_0}\frac{\delta\gamma_0}{\gamma_0}.
\end{equation}
During the interaction with the scattering laser, the electron energy is converted into radiation. The total number of single photon - electron collisions per electron ($N_{col}$) can be estimated knowing the laser intensity, frequency and duration, and using the Compton cross section and the Poynting flux in the average electron rest frame \cite{RPastro} (quantities in electron rest frame are marked with a star) 
$dN_{col}/dt^*=\sigma_C(c/8\pi)(E^{*2}+B^{*2})/(\hbar \omega_0^*)$.
In the laboratory frame, this yields $dN_{col}/dt=e^2 a_0^2(t)\omega_0/(3 c \hbar)$, which for the $\sin^2$ temporal laser envelope gives: 
\begin{equation}
N_{col}=\frac{e^2a_0^2\omega_0}{8c\hbar}\mathrm{\tau_{0}}=1.72\times10^{-3}a_0^2\tau_\mathrm{0}[\mathrm{fs}]\left(\frac{1\mu\mathrm{m}}{\lambda_0} \right),
\end{equation}
where $\lambda_0$ is the wavelength of the scattering laser. The total number of collisions does not depend on the electron initial energy, but only on the photon density (laser intensity). If the number of electrons in the LWFA beam is $N_e$, the total number of backscattered photons during the interaction with the laser is simply given by $N_\gamma=N_eN_{col}$. The number of self-injected electrons in a matched LWFA is $N_e\simeq (1/30) (2 \sqrt{a_{LD}})^3 (1/k_p r_0)$, where $k_p=\omega_p/c$ is the plasma wavenumber. Hence, the total number of photons is approximately given by $N_{\gamma}=2.59\times10^{4}a_0^2\tau_\mathrm{0} [\mathrm{fs}] a_{LD}^{3/2}\left(\lambda_p/\lambda_0 \right)$,
where $\lambda_p$ is the plasma wavelength. This can be written in terms of laser intensity yielding
$
N_{\gamma}=1.497\times10^{12}I_0 \left[ \mathrm{ 10^{22}W/cm^2} \right]\lambda_0[\mathrm{\mu m}]\tau_0[10 \mathrm{fs}] \times I_{LD}^{3/4}\left[\mathrm{10^{22} W/cm^2}\right]\lambda_{LD}^{3/2}[\mathrm{\mu m}]\lambda_p[\mathrm{\mu m}] 
$, where $\lambda_{LD}$ is the wavelength of the driver. For relativistic electrons, the radiation is confined within a narrow angle that scales with $1/\gamma$ around the propagation direction. In our setup, the counter-propagating laser is linearly polarised in the $x_3$ direction, and propagates in the negative $x_1$ direction. The electrons wiggle in the laser polarisation plane, and the maximum angle of the electron momentum with respect to the initial propagation direction is $p_3/p_1\approx a_0/\gamma$. If $a_0>1$,  and using the LWFA output electron energy, the maximum angle of the radiation of a single electron is given by
$
\theta_{rad}=(3a_0/2a_{LD}) \left(\omega_p/\omega_{LD} \right)^2.
$

For a head-on collision with $a_0\gg1$, the photons are radiated in the nonlinear regime, with the first harmonic fundamental frequency on-axis given by 
$\omega_R\simeq4\omega_0\gamma^2/\alpha$, where $\alpha=1+a_0^2/2$ for a linearly polarised wave and $\alpha=1+a_0^2$ for a circularly polarised wave. Classically, this can be seen as a double Doppler shift of the laser photon due to the parallel component of the electron motion. In QED, $\alpha$ comes from the electron relativistic mass shift \cite{NikishovRitus, kogaPOP}. Hence, the fundamental frequency of the emitted on-axis radiation during the interaction of the self-injected bunch and the scattering pulse is on the order of $\omega_R\simeq (16/9) (\omega_{LD}/\omega_p)^4 a_{LD}^2\omega_0/(1+a_0^2/2)$, corresponding to $\sim 10^{11}$ photons with energies on the order of 44 keV for the scattering of GeV-class electron bunches with 30fs long laser pulses with the intensity of $\mathrm{10^{21} W/cm^2}$. Ideally, the goal would be to observe radiation reaction signatures in the radiated spectrum. However, this may not be possible since the main effect of radiation reaction is the slowdown of the colliding electrons which in turn smears out the possible radiation signatures. In fact, as shown by eqs. \eqref{radP} - \eqref{efinal}, in the aforementioned case the electron bunch energy decreases by more than 40\%.

 \begin{figure}
 \begin{center}
  \includegraphics[width=25em]{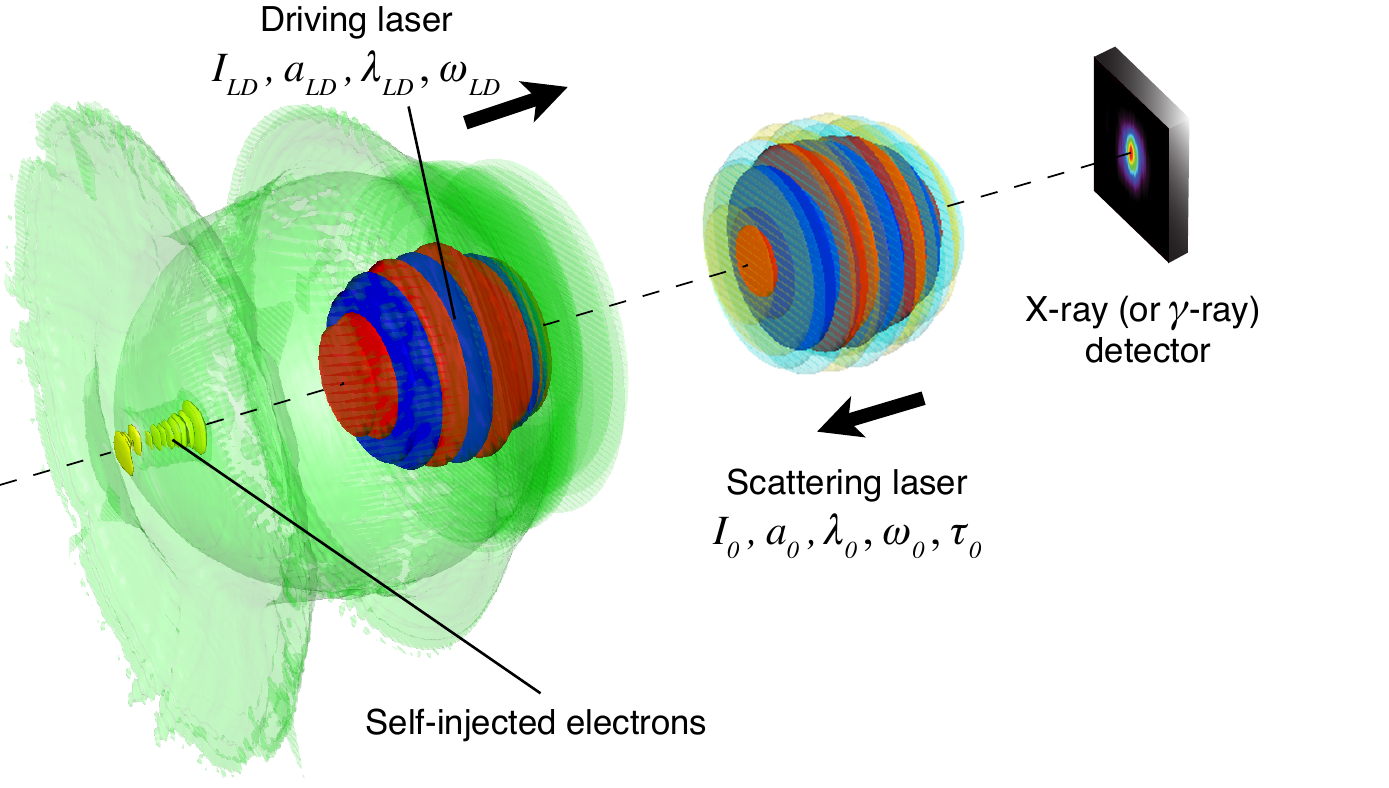}
  \end{center}
 \caption{ \label{setup} \textbf{All-optical radiation reaction configuration}. A moderate intensity laser is used to generate the laser wakefield where the electrons are self-injected and accelerated. A counter-propagating ultra high intensity laser pulse collides head-on with the energetic electron bunch in vacuum after it leaves the plasma.       }
 \end{figure}

We now explore this configuration resorting to 3D full-scale PIC simulations with code OSIRIS, over a wide range of parameters encompassing current and near future laser technology. OSIRIS \cite{OSIRIS} is a fully-relativistic PIC code, where for the purpose of this study the standard Lorentz force-based particle pusher was replaced by a LL pusher \cite{myAPS}. Diffraction-limited pulses are considered, with longitudinal profiles given by $10\tau^3-15\tau^4+6\tau^5$, with $\tau=\sqrt{2}t/\tau_0$, and $\tau_0$ being the pulse duration at FWHM. The transverse profile of the lasers is Gaussian with the spot size defined as FWHM in the fields. Laser parameters for different simulations are summarized in Table 1. Lasers \textbf{a}, \textbf{b}, \textbf{c} were used to obtain 0.5, 1 and 1.5 GeV electrons in the LWFA, which after leaving the plasma interacted with the more intense scaterring lasers A-E in vacuum. When simulating LWFA corresponding to the laser \textbf{a} the plasma slab was 3.4 mm long, with an electron density of $2.8 \times 10^{18} \mathrm{cm^{ -3}}$.  For each of these runs, the simulation box size was $\mathrm{76.4\  \mu m\times101.9 \ \mu m \times 101.9\  \mu m}$, with $\mathrm{2400\times160\times160}$ cells, and the total simulation time was 11.6 ps. Each simulation had $1.3\times 10^8$ particles pushed for $1.16\times10^5$ timesteps. The mean output energy of the LWFA electron beam was 0.51 GeV. The LWFA corresponding to the laser \textbf{b} in the second column of Table 1 had plasma conditions similar to the laser \textbf{a}. For these runs, the simulation box dimensions were $\mathrm{95.5\  \mu m\times 152.8 \ \mu m \times 152.8\  \mu m}$, the number of cells was $\mathrm{3000\times240\times240}$, and the total time was 12.7 ps. Each simulation had $3.5\times 10^8$ particles pushed for $1.25\times10^5$ timesteps. The mean output energy of the electron beam was 0.93 GeV. 
 The LWFA corresponding to the laser \textbf{c} had a plasma slab 7.66 mm long, with density $2.107 \times 10^{18} \mathrm{cm^{ -3}}$, leading to 1.5 GeV - class electron bunches. For these runs, the simulation box size was $\mathrm{132.1\  \mu m\times 175.1 \ \mu m \times 175.1\  \mu m}$, the number of cells used was $\mathrm{4150\times240\times240}$, and the total time was 26.3 ps. Each simulation had $4.8\times 10^8$ particles pushed for $2.6\times10^5$ timesteps. The mean output energy of the electron beam was 1.55 GeV.

\begin{table*}\begin{tabular}{lrrrrp{3mm}rp{3mm}rp{1cm}crp{3mm}rp{3mm}rp{3mm}rp{3mm}r} 
&&&\multicolumn{6}{c}{Driving laser}&&&\multicolumn{9}{c}{Scattering laser}\\
\cline{5-9} \cline{12-20}
&&&&\textbf{a}&&\textbf{b}&&\textbf{c}&&&\textbf{A}&&\textbf{B}&&\textbf{C}&&\textbf{D}&&\textbf{E}\\
\hline
$a_0$ &&&&4&&8&&9&&&8.55&&17.1&&27.0&&54.0&&85.5\\
Spot ($\mu$m)&&&& 13 && 18 && 22 &&& 10 && 10 && 10 && 10 && 10 \\
Duration (fs)&&&& 42.4 && 60.0 && 73.3 &&& 26.5 && 26.5 && 26.5 && 26.5 && 26.5\\ 
Power (PW) &&&& 0.044 && 0.349 && 0.658 &&& 0.123 && 0.491 && 1.23 && 4.91 &&  12.3 \\
Energy (J) &&&& 1.855 && 21 && 48.2 &&& 4 && 16.4 && 40.8 && 164 && 410\\
Intensity ($10^{20}$ W/cm$^2$)&&&& 0.22 && 0.88 && 1.1 &&& 1 && 4 && 10 && 40 && 100  \\
$\omega_{LD}/\omega_p$&&&& 20 && 20 && 23 &&& &&  &&  &&  &&  \\
\hline 
\end{tabular} 
\caption{\textbf{Laser parameters for the parameter scan.} LWFA with lasers \textbf{a}, \textbf{b} and \textbf{c} and plasma slabs with density of the order $10^{18}\mathrm{cm^{-3}}$ are simulated in matched conditions for the blowout regime \cite{LWFAscailing}, leading to acceleration of 0.5, 1 and 1.5 GeV electron bunches respectively. As they leave the plasma, electron bunches are scattered by counter-propagating lasers A, B, C, D or E. All the lasers in the setup have wavelength of $1 \mu m$.}
 \end{table*}

 \begin{figure} 
 \begin{center}
  \includegraphics[width=25em]{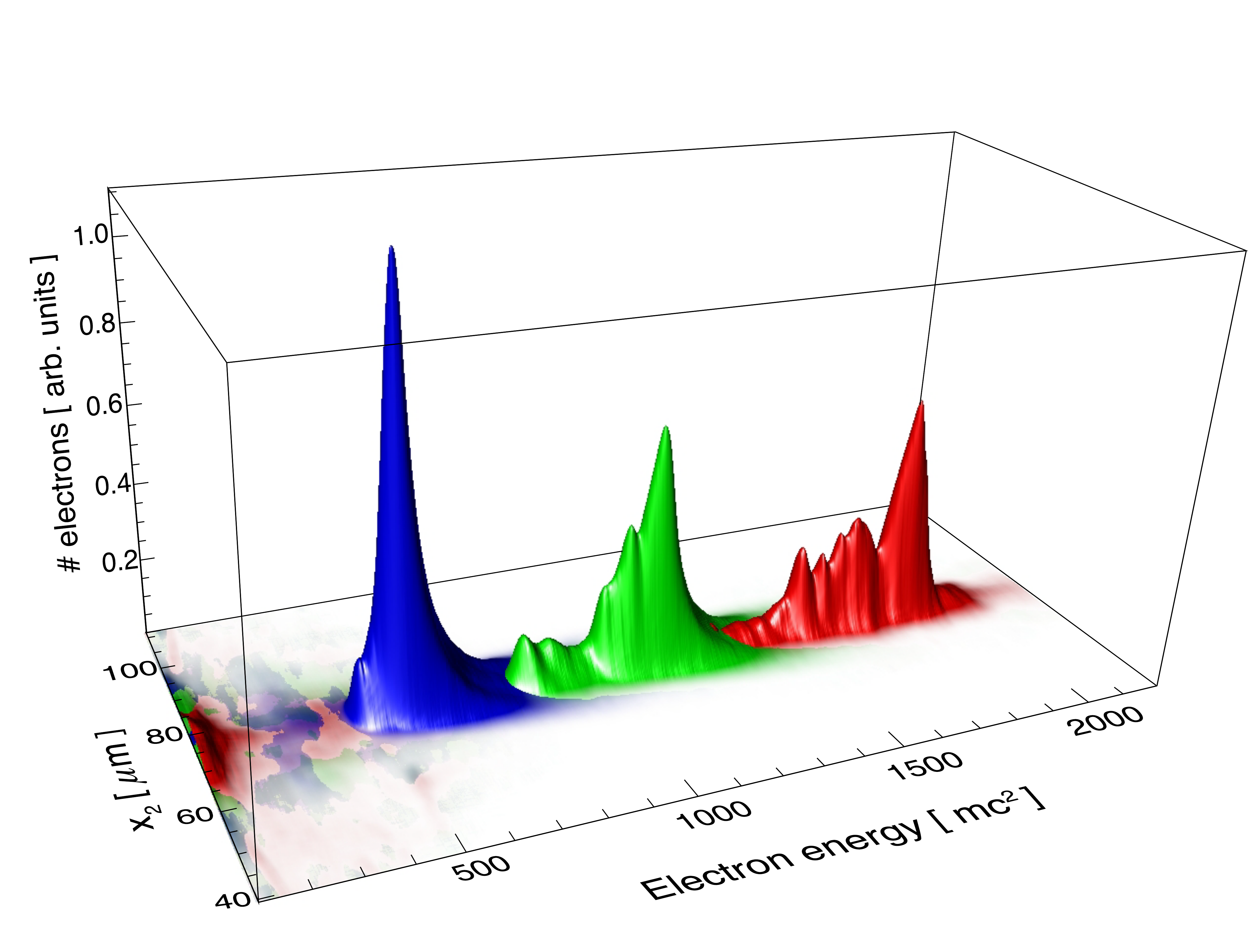}
 \end{center}
 \caption{ \label{spect3d} \textbf{LWFA electron beam profile initially, and after interaction with two different lasers.} The initial beam profile given in red, and after interacting with a $10^{21} \mathrm{ W / cm^2}$ laser (in green) or after interacting with a  $4\times10^{21} \mathrm{ W / cm^2}$ (in blue). While losing energy, the beam profile becomes more uniform and the energy spread decreases, according to the Eq.\eqref{enspread}. The peak positions are in agreement with the theoretical predictions of Eq. \eqref{efinal}. }
 \end{figure}

\begin{figure}
 \begin{center}
  \includegraphics[width=26em]{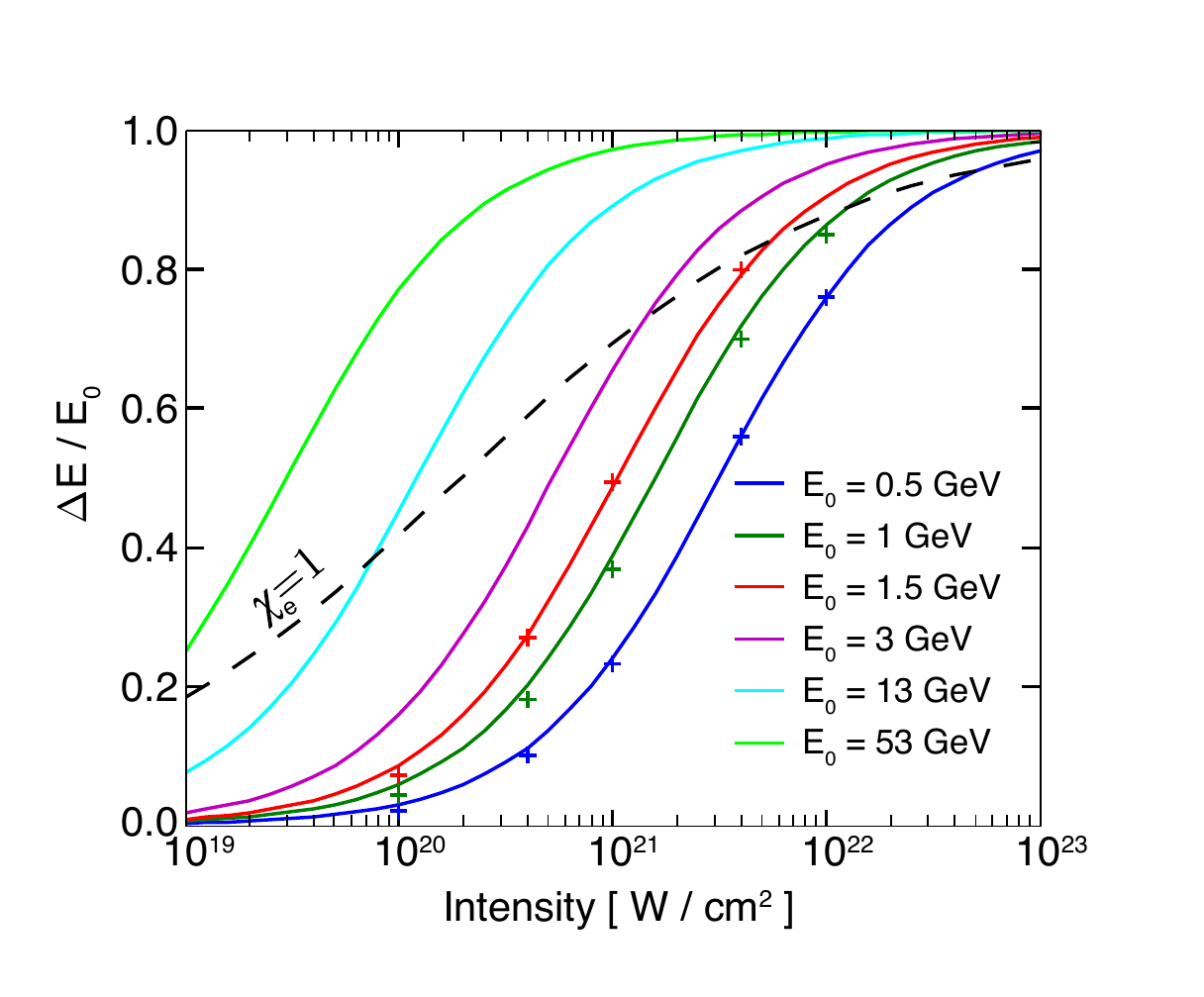}
  \end{center}
 \caption{ \label{spect} \textbf{Electron beam energy loss.} Parameter scan based on \emph{ab initio} full scale PIC simulations for different experimental conditions that correspond to 0.5 GeV, 1 GeV and 1.5 GeV - class LWFA electron beams, coupled with a scattering laser with intensity in the range $\mathrm{10^{20} - 10^{22} W / cm^2}$. Curves represent the theoretical prediction of Eq. \eqref{efinal}, and each cross represents one simulation result. For reference, the curves for higher energy electron beams from ref. \cite{Samuel_nature} are also given; the dashed black line corresponds to the value $\chi_e=1$, above which we can expect to produce electron-positron pairs and quantum effects can start to play a significant role in the electron dynamics.   }
 \end{figure}

Quasi-mono-energetic electron bunches with peak energies ranging from 0.5 - 1.5 GeV are generated in the LWFA stage. The scattering laser, with a 10 $\mu$m spot-size, is much wider than the transverse width of the LWFA electron bunch, on the order of 2 microns, thus suggesting that the electron beam is fully scattered by the laser field. Therefore Eqs.  \eqref{radP} - \eqref{efinal} can be employed to estimate the electron energy loss in the interaction. Large energy losses (40\% for 1 GeV electrons colliding with a $10^{21} \mathrm{W/cm^2}$ laser) can be easily measured in an experiment, even if the electron bunch is not quasi-monoenergetic (see Fig. \ref{spect3d}). Excellent agreement between analytical and numerical results is obtained, as shown in Fig. \ref{spect}.  The parameter $\chi_e$ represents the ratio between the maximal laser electric field amplitude in the electron rest frame and the critical Schwinger field \cite{Schwinger}, and the curve corresponding to $\chi_e=1$ marks the theoretical transition between the classical and the quantum radiation reaction dominated regime. This transition has not been explored experimentally up to date, and Fig. \ref{spect} shows that the near-future laser technology with intensities above $10^{22} \mathrm{W/cm^2}$, in combination with multi - GeV electron bunches generated with even more modest laser intensities and energies, will open the path for its experimental verification and exploration of signatures. 

\begin{figure}
 \begin{center}
  \includegraphics[width=28em]{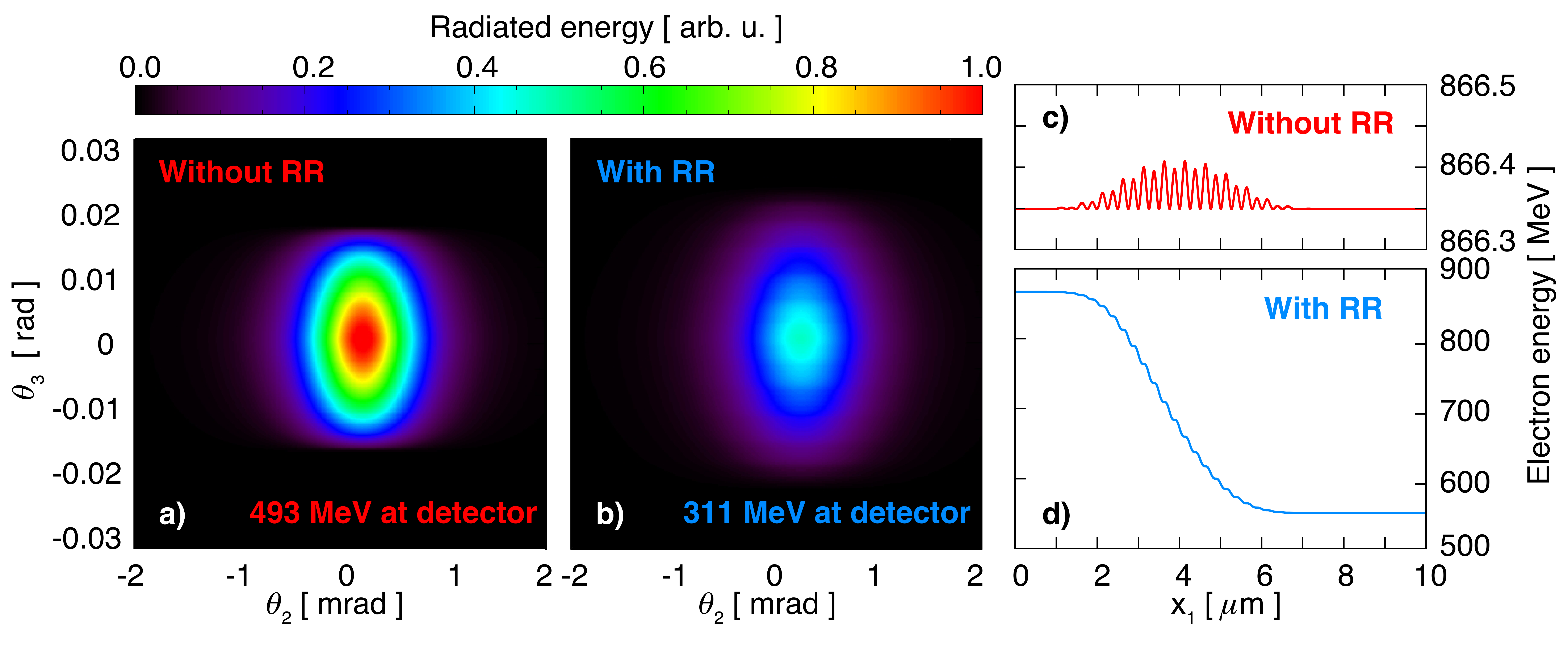}
  \end{center}
 \caption{ \label{one_part} \textbf{Single electron radiation.} The radiated energy captured on a virtual detector a) without and b) with radiation reaction; evolution of the electron energy vs. longitudinal 
position, without c) and with d) the radiation reaction. When accounting for radiation reaction, the particle loses 315.4 MeV in the interaction, which is consistent with the energy captured on the detector b). Without radiation reaction, the particle does not lose any energy, but appears to have radiated over 50\% of its total energy to the detector. 
  }
 \end{figure}

We can isolate a single test electron and numerically integrate its trajectory in the laser field with and without accounting for radiation reaction in otherwise identical conditions (see Fig \ref{one_part}). A post-processing diagnostic JRAD \cite{Joana} is then used to deposit the radiated fields on a virtual detector in both cases. JRAD uses the phase-space trajectory  of the electrons to calculate the total energy received in each pixel. Integration over time and over the surface of the detector gives the total radiated energy received on the detector for the whole interaction time. The detector captures 493 MeV when radiation reaction is not included in the electron motion (Fig. \ref{one_part}a), and 311 MeV if the radiation reaction is accounted for (Fig. \ref{one_part}b). The total energy loss of the test electron due to radiation reaction is 315 MeV. This means that the purely classical calculations, which ignore the radiation reaction, overestimate the total emitted radiation in this scenario and lead to results that are inconsistent with energy conservation laws. As expected, the energy lost by the electron when accounting for radiation reaction is fully consistent with the radiated photon energy captured on the detector. 

\begin{figure}
 \begin{center}
  \includegraphics[width=26em]{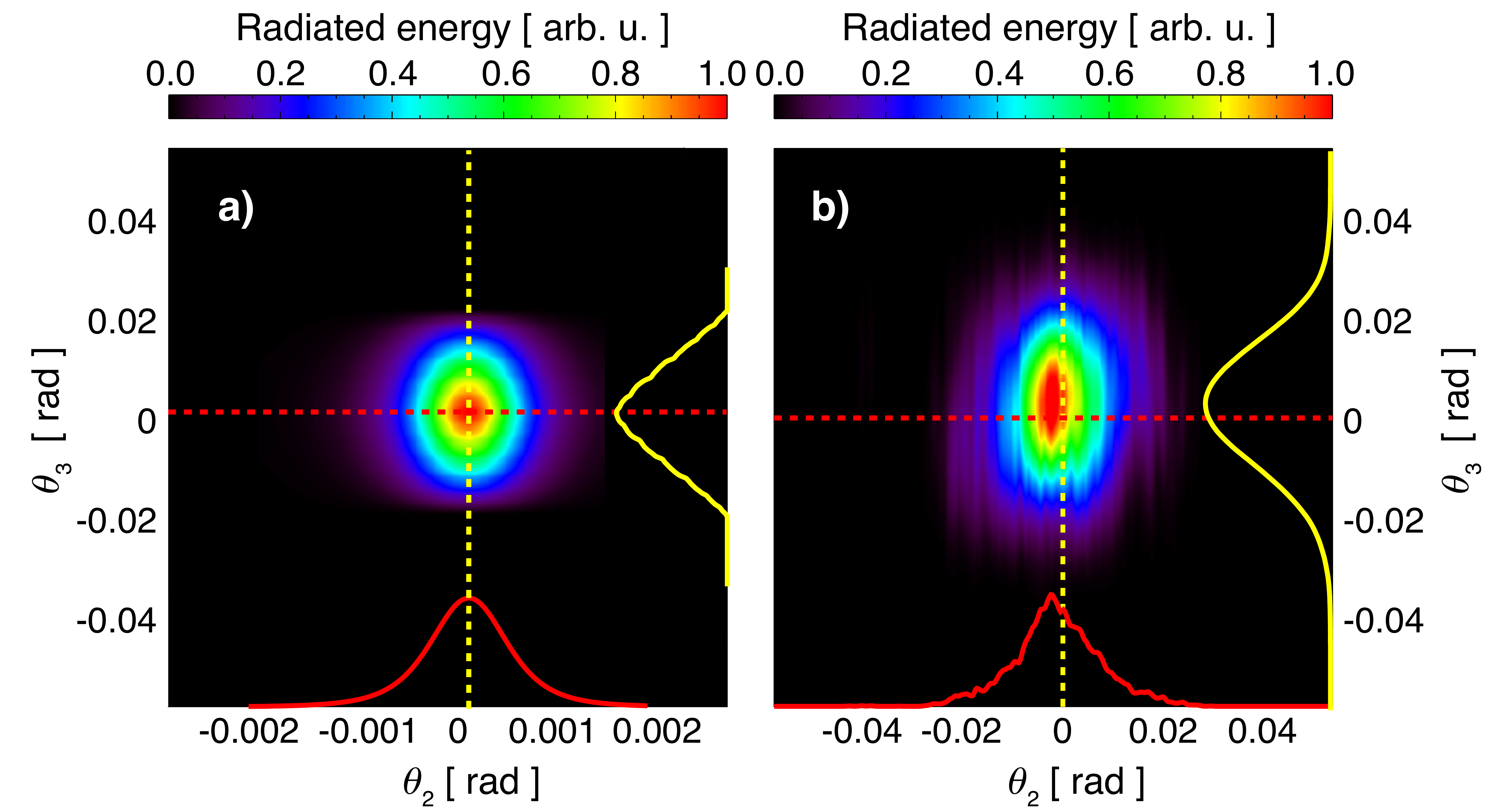}
  \end{center}
 \caption{ \label{rad} \textbf{Emitted radiation from PIC simuations.} \textbf{a)} Single electron radiated energy captured in a virtual detector located 4 cm from the interaction point.  The electron relativistic factor decreases from $\gamma\approx1700$ to $\gamma\approx1100$. \textbf{b)} Total radiated energy of the electron beam in the same detector; over 99\% of the radiated energy is captured. The horizontal scale is 20 times larger than in a), so here a single particle contribution appears as a thin vertical line. The electron beam divergence in $x_2$ is an order of magnitude higher than $1/\gamma$, so the maximum radiation angle is determined by the beam divergence.  }
 \end{figure}

The total radiated energy of a single electron from OSIRIS simulations captured in a virtual detector located 4 cm from the interaction region is shown in Fig. \ref{rad}a. Single electron radiated pattern in Fig \ref{rad}a is similar to that of test particle integrated with few orders of magnitude higher resolution in Fig \ref{one_part}. Figure \ref{rad}b shows the radiation on the same detector for the LWFA electron beam, represented by 1\% random sample, where the energy captured on the detector is above 99\% of the energy lost by the electrons.

Therefore, analytical estimates, full-scale 3D PIC simulations including radiation reaction and a post-processing radiation diagnostic all consistently predict the same electron energy loss, which is measurable in present-day laboratory conditions, and thus provides a direct signature for radiation reaction and a path to explore the classical to quantum radiation reaction transition.


\bibliography{RadiationDamping.bib}

\begin{thebibliography}{10}%
\makeatletter
\providecommand \@ifxundefined [1]{%
 \ifx #1\undefined \expandafter \@firstoftwo
 \else \expandafter \@secondoftwo
\fi
}%
\providecommand \@ifnum [1]{%
 \ifnum #1\expandafter \@firstoftwo
 \else \expandafter \@secondoftwo
\fi
}%
\providecommand \enquote [1]{``#1''}%
\providecommand \bibnamefont  [1]{#1}%
\providecommand \bibfnamefont [1]{#1}%
\providecommand \citenamefont [1]{#1}%
\providecommand\href[0]{\@sanitize\@href}%
\providecommand\@href[1]{\endgroup\@@startlink{#1}\endgroup\@@href}%
\providecommand\@@href[1]{#1\@@endlink}%
\providecommand \@sanitize [0]{\begingroup\catcode`\&12\catcode`\#12\relax}%
\@ifxundefined \pdfoutput {\@firstoftwo}{%
 \@ifnum{\z@=\pdfoutput}{\@firstoftwo}{\@secondoftwo}%
}{%
 \providecommand\@@startlink[1]{\leavevmode\special{html:<a href="#1">}}%
 \providecommand\@@endlink[0]{\special{html:</a>}}%
}{%
 \providecommand\@@startlink[1]{%
  \leavevmode
  \pdfstartlink
   attr{/Border[0 0 1 ]/H/I/C[0 1 1]}%
   user{/Subtype/Link/A<</Type/Action/S/URI/URI(#1)>>}%
  \relax
 }%
 \providecommand\@@endlink[0]{\pdfendlink}%
}%
\providecommand \url  [0]{\begingroup\@sanitize \@url }%
\providecommand \@url [1]{\endgroup\@href {#1}{\urlprefix}}%
\providecommand \urlprefix [0]{URL }%
\providecommand \Eprint[0]{\href }%
\@ifxundefined \urlstyle {%
  \providecommand \doi [1]{doi:\discretionary{}{}{}#1}%
}{%
  \providecommand \doi [0]{doi:\discretionary{}{}{}\begingroup
  \urlstyle{rm}\Url }%
}%
\providecommand \doibase [0]{http://dx.doi.org/}%
\providecommand \Doi[1]{\href{\doibase#1}}%
\providecommand \bibAnnote [3]{%
  \BibitemShut{#1}%
  \begin{quotation}\noindent
    \textsc{Key:}\ #2\\\textsc{Annotation:}\ #3%
  \end{quotation}%
}%
\providecommand \bibAnnoteFile [2]{%
  \IfFileExists{#2}{\bibAnnote {#1} {#2} {\input{#2}}}{}%
}%
\providecommand \typeout [0]{\immediate \write \m@ne }%
\providecommand \selectlanguage [0]{\@gobble}%
\providecommand \bibinfo [0]{\@secondoftwo}%
\providecommand \bibfield [0]{\@secondoftwo}%
\providecommand \translation [1]{[#1]}%
\providecommand \BibitemOpen[0]{}%
\providecommand \bibitemStop [0]{}%
\providecommand \bibitemNoStop [0]{.\EOS\space}%
\providecommand \EOS [0]{\spacefactor3000\relax}%
\providecommand \BibitemShut [1]{\csname bibitem#1\endcsname}%
\bibitem{LAD}%
  \BibitemOpen
  \bibfield{author}{%
  \bibinfo {author} {\bibfnamefont{P.~A.~M.}\ \bibnamefont{Dirac}},\ }%
  \bibfield{journal}{%
  \Doi{10.1098/rspa.1938.0124}{\bibinfo {journal} {Proc. R. Soc. Lond. A}}\ }%
  \textbf{\bibinfo {volume} {167}},\ \bibinfo {pages} {148} (\bibinfo {year}
  {1938})%
  \bibAnnoteFile{NoStop}{LAD}%
\bibitem{LLbook}%
  \BibitemOpen
  \bibfield{author}{%
  \bibinfo {author} {\bibfnamefont{L.~D.}\ \bibnamefont{{Landau}}}\ and\
  \bibinfo {author} {\bibfnamefont{E.~M.}\ \bibnamefont{{Lifshitz}}},\ }%
  \emph{\bibinfo {title} {The Classical Theory of Fields}}\ (\bibinfo
  {publisher} {Butterworth Heinemann},\ \bibinfo {year} {1975})%
  \bibAnnoteFile{NoStop}{LLbook}%
\bibitem{model1bell}%
  \BibitemOpen
  \bibfield{author}{%
  \bibinfo {author} {\bibfnamefont{A.~R.}\ \bibnamefont{Bell}}\ and\ \bibinfo
  {author} {\bibfnamefont{J.~G.}\ \bibnamefont{Kirk}},\ }%
  \bibfield{journal}{%
  \Doi{10.1103/PhysRevLett.101.200403}{\bibinfo {journal} {Phys. Rev. Lett.}}\
  }%
  \textbf{\bibinfo {volume} {101}},\ \bibinfo {pages} {200403} (\bibinfo
  {month} {Nov}\ \bibinfo {year} {2008})%
  \bibAnnoteFile{NoStop}{model1bell}%
\bibitem{model3sokolov}%
  \BibitemOpen
  \bibfield{author}{%
  \bibinfo {author} {\bibfnamefont{I.~V.}\ \bibnamefont{Sokolov}}, \bibinfo
  {author} {\bibfnamefont{N.~M.}\ \bibnamefont{Naumova}}, \bibinfo {author}
  {\bibfnamefont{J.~A.}\ \bibnamefont{Nees}}, \bibinfo {author}
  {\bibfnamefont{G.~A.}\ \bibnamefont{Mourou}},\ and\ \bibinfo {author}
  {\bibfnamefont{V.~P.}\ \bibnamefont{Yanovsky}},\ }%
  \bibfield{journal}{%
  \Doi{10.1063/1.3236748}{\bibinfo {journal} {Physics of Plasmas}}\ }%
  \textbf{\bibinfo {volume} {16}},\ \bibinfo {eid} {093115} (\bibinfo {year}
  {2009})%
  \bibAnnoteFile{NoStop}{model3sokolov}%
\bibitem{spohn}%
  \BibitemOpen
  \bibfield{author}{%
  \bibinfo {author} {\bibfnamefont{H.}~\bibnamefont{Spohn}},\ }%
  \bibfield{journal}{%
  \Doi{10.1209/epl/i2000-00268-x}{\bibinfo {journal} {Europhys. Lett.}}\ }%
  \textbf{\bibinfo {volume} {50}},\ \bibinfo {pages} {287} (\bibinfo {month}
  {May}\ \bibinfo {year} {2000})%
  \bibAnnoteFile{NoStop}{spohn}%
\bibitem{rorlich_1}%
  \BibitemOpen
  \bibfield{author}{%
  \bibinfo {author} {\bibfnamefont{F.}~\bibnamefont{Rohrlich}},\ }%
  \bibfield{journal}{%
  \Doi{10.1103/PhysRevE.77.046609}{\bibinfo {journal} {Phys. Rev. E}}\ }%
  \textbf{\bibinfo {volume} {77}},\ \bibinfo {pages} {046609} (\bibinfo {month}
  {Apr}\ \bibinfo {year} {2008})%
  \bibAnnoteFile{NoStop}{rorlich_1}%
\bibitem{Ruhl_threshold}%
  \BibitemOpen
  \bibfield{author}{%
  \bibinfo {author} {\bibfnamefont{Y.}~\bibnamefont{Hadad}}, \bibinfo {author}
  {\bibfnamefont{L.}~\bibnamefont{Labun}}, \bibinfo {author}
  {\bibfnamefont{J.}~\bibnamefont{Rafelski}}, \bibinfo {author}
  {\bibfnamefont{N.}~\bibnamefont{Elkina}}, \bibinfo {author}
  {\bibfnamefont{C.}~\bibnamefont{Klier}},\ and\ \bibinfo {author}
  {\bibfnamefont{H.}~\bibnamefont{Ruhl}},\ }%
  \bibfield{journal}{%
  \Doi{10.1103/PhysRevD.82.096012}{\bibinfo {journal} {Phys. Rev. D}}\ }%
  \textbf{\bibinfo {volume} {82}},\ \bibinfo {pages} {096012} (\bibinfo {month}
  {Nov}\ \bibinfo {year} {2010})%
  \bibAnnoteFile{NoStop}{Ruhl_threshold}%
\bibitem{capturePiazza}%
  \BibitemOpen
  \bibfield{author}{%
  \bibinfo {author} {\bibfnamefont{A.}~\bibnamefont{Di~Piazza}}, \bibinfo
  {author} {\bibfnamefont{K.~Z.}\ \bibnamefont{Hatsagortsyan}},\ and\ \bibinfo
  {author} {\bibfnamefont{C.~H.}\ \bibnamefont{Keitel}},\ }%
  \bibfield{journal}{%
  \Doi{10.1103/PhysRevLett.102.254802}{\bibinfo {journal} {Phys. Rev. Lett.}}\
  }%
  \textbf{\bibinfo {volume} {102}},\ \bibinfo {pages} {254802} (\bibinfo
  {month} {Jun}\ \bibinfo {year} {2009})%
  \bibAnnoteFile{NoStop}{capturePiazza}%
\bibitem{ThomasRR}%
  \BibitemOpen
  \bibfield{author}{%
  \bibinfo {author} {\bibfnamefont{A.~G.~R.}\ \bibnamefont{Thomas}}, \bibinfo
  {author} {\bibfnamefont{C.~P.}\ \bibnamefont{Ridgers}}, \bibinfo {author}
  {\bibfnamefont{S.~S.}\ \bibnamefont{Bulanov}}, \bibinfo {author}
  {\bibfnamefont{B.~J.}\ \bibnamefont{Griffin}},\ and\ \bibinfo {author}
  {\bibfnamefont{S.~P.~D.}\ \bibnamefont{Mangles}},\ }%
  \bibfield{journal}{%
  \Doi{10.1103/PhysRevX.2.041004}{\bibinfo {journal} {Phys. Rev. X}}\ }%
  \textbf{\bibinfo {volume} {2}},\ \bibinfo {pages} {041004} (\bibinfo {month}
  {Oct}\ \bibinfo {year} {2012}),\
  \url{http://link.aps.org/doi/10.1103/PhysRevX.2.041004}%
  \bibAnnoteFile{NoStop}{ThomasRR}%
\bibitem{Lwfa_gev_1}%
  \BibitemOpen
  \bibfield{author}{%
  \bibinfo {author} {\bibfnamefont{W.~P.}\ \bibnamefont{Leemans}}, \bibinfo
  {author} {\bibfnamefont{B.}~\bibnamefont{Nagler}}, \bibinfo {author}
  {\bibfnamefont{A.~J.}\ \bibnamefont{Gonsalves}}, \bibinfo {author}
  {\bibfnamefont{C.}~\bibnamefont{Toth}}, \bibinfo {author}
  {\bibfnamefont{K.}~\bibnamefont{Nakamura}}, \bibinfo {author}
  {\bibfnamefont{C.~G.~R.}\ \bibnamefont{Geddes}}, \bibinfo {author}
  {\bibfnamefont{E.}~\bibnamefont{Esarey}}, \bibinfo {author}
  {\bibfnamefont{C.~B.}\ \bibnamefont{Schroeder}},\ and\ \bibinfo {author}
  {\bibfnamefont{S.~M.}\ \bibnamefont{Hooker}},\ }%
  \bibfield{journal}{%
  \Doi{10.1038/nphys418}{\bibinfo {journal} {Nature Phys.}}\ }%
  \textbf{\bibinfo {volume} {2}},\ \bibinfo {pages} {696} (\bibinfo {month}
  {Sept}\ \bibinfo {year} {2006})%
  \bibAnnoteFile{NoStop}{Lwfa_gev_1}%
\bibitem{Lwfa_gev_2}%
  \BibitemOpen
  \bibfield{author}{%
  \bibinfo {author} {\bibfnamefont{S.}~\bibnamefont{Kneip}}, \bibinfo {author}
  {\bibfnamefont{S.~R.}\ \bibnamefont{Nagel}}, \bibinfo {author}
  {\bibfnamefont{S.~F.}\ \bibnamefont{Martins}}, \bibinfo {author}
  {\bibfnamefont{S.~P.~D.}\ \bibnamefont{Mangles}}, \bibinfo {author}
  {\bibfnamefont{C.}~\bibnamefont{Bellei}}, \bibinfo {author}
  {\bibfnamefont{O.}~\bibnamefont{Chekhlov}}, \bibinfo {author}
  {\bibfnamefont{R.~J.}\ \bibnamefont{Clarke}}, \bibinfo {author}
  {\bibfnamefont{N.}~\bibnamefont{Delerue}}, \bibinfo {author}
  {\bibfnamefont{E.~J.}\ \bibnamefont{Divall}}, \bibinfo {author}
  {\bibfnamefont{G.}~\bibnamefont{Doucas}}, \bibinfo {author}
  {\bibfnamefont{K.}~\bibnamefont{Ertel}}, \bibinfo {author}
  {\bibfnamefont{F.}~\bibnamefont{Fiuza}}, \bibinfo {author}
  {\bibfnamefont{R.}~\bibnamefont{Fonseca}}, \bibinfo {author}
  {\bibfnamefont{P.}~\bibnamefont{Foster}}, \bibinfo {author}
  {\bibfnamefont{S.~J.}\ \bibnamefont{Hawkes}}, \bibinfo {author}
  {\bibfnamefont{C.~J.}\ \bibnamefont{Hooker}}, \bibinfo {author}
  {\bibfnamefont{K.}~\bibnamefont{Krushelnick}}, \bibinfo {author}
  {\bibfnamefont{W.~B.}\ \bibnamefont{Mori}}, \bibinfo {author}
  {\bibfnamefont{C.~A.~J.}\ \bibnamefont{Palmer}}, \bibinfo {author}
  {\bibfnamefont{K.~T.}\ \bibnamefont{Phuoc}}, \bibinfo {author}
  {\bibfnamefont{P.~P.}\ \bibnamefont{Rajeev}}, \bibinfo {author}
  {\bibfnamefont{J.}~\bibnamefont{Schreiber}}, \bibinfo {author}
  {\bibfnamefont{M.~J.~V.}\ \bibnamefont{Streeter}}, \bibinfo {author}
  {\bibfnamefont{D.}~\bibnamefont{Urner}}, \bibinfo {author}
  {\bibfnamefont{J.}~\bibnamefont{Vieira}}, \bibinfo {author}
  {\bibfnamefont{L.~O.}\ \bibnamefont{Silva}},\ and\ \bibinfo {author}
  {\bibfnamefont{Z.}~\bibnamefont{Najmudin}},\ }%
  \bibfield{journal}{%
  \Doi{10.1103/PhysRevLett.103.035002}{\bibinfo {journal} {Phys. Rev. Lett.}}\
  }%
  \textbf{\bibinfo {volume} {103}},\ \bibinfo {pages} {035002} (\bibinfo
  {month} {Jul}\ \bibinfo {year} {2009})%
  \bibAnnoteFile{NoStop}{Lwfa_gev_2}%
\bibitem{Lwfa_gev_3}%
  \BibitemOpen
  \bibfield{author}{%
  \bibinfo {author} {\bibfnamefont{D.~H.}\ \bibnamefont{Froula}}, \bibinfo
  {author} {\bibfnamefont{C.~E.}\ \bibnamefont{Clayton}}, \bibinfo {author}
  {\bibfnamefont{T.}~\bibnamefont{D\"oppner}}, \bibinfo {author}
  {\bibfnamefont{K.~A.}\ \bibnamefont{Marsh}}, \bibinfo {author}
  {\bibfnamefont{C.~P.~J.}\ \bibnamefont{Barty}}, \bibinfo {author}
  {\bibfnamefont{L.}~\bibnamefont{Divol}}, \bibinfo {author}
  {\bibfnamefont{R.~A.}\ \bibnamefont{Fonseca}}, \bibinfo {author}
  {\bibfnamefont{S.~H.}\ \bibnamefont{Glenzer}}, \bibinfo {author}
  {\bibfnamefont{C.}~\bibnamefont{Joshi}}, \bibinfo {author}
  {\bibfnamefont{W.}~\bibnamefont{Lu}}, \bibinfo {author}
  {\bibfnamefont{S.~F.}\ \bibnamefont{Martins}}, \bibinfo {author}
  {\bibfnamefont{P.}~\bibnamefont{Michel}}, \bibinfo {author}
  {\bibfnamefont{W.~B.}\ \bibnamefont{Mori}}, \bibinfo {author}
  {\bibfnamefont{J.~P.}\ \bibnamefont{Palastro}}, \bibinfo {author}
  {\bibfnamefont{B.~B.}\ \bibnamefont{Pollock}}, \bibinfo {author}
  {\bibfnamefont{A.}~\bibnamefont{Pak}}, \bibinfo {author}
  {\bibfnamefont{J.~E.}\ \bibnamefont{Ralph}}, \bibinfo {author}
  {\bibfnamefont{J.~S.}\ \bibnamefont{Ross}}, \bibinfo {author}
  {\bibfnamefont{C.~W.}\ \bibnamefont{Siders}}, \bibinfo {author}
  {\bibfnamefont{L.~O.}\ \bibnamefont{Silva}},\ and\ \bibinfo {author}
  {\bibfnamefont{T.}~\bibnamefont{Wang}},\ }%
  \bibfield{journal}{%
  \Doi{10.1103/PhysRevLett.103.215006}{\bibinfo {journal} {Phys. Rev. Lett.}}\
  }%
  \textbf{\bibinfo {volume} {103}},\ \bibinfo {pages} {215006} (\bibinfo
  {month} {Nov}\ \bibinfo {year} {2009})%
  \bibAnnoteFile{NoStop}{Lwfa_gev_3}%
\bibitem{MalkaNature}%
  \BibitemOpen
  \bibfield{author}{%
  \bibinfo {author} {\bibfnamefont{K.}~\bibnamefont{Ta~Phuoc}}, \bibinfo
  {author} {\bibfnamefont{S.}~\bibnamefont{Corde}}, \bibinfo {author}
  {\bibfnamefont{C.}~\bibnamefont{Thaury}}, \bibinfo {author}
  {\bibfnamefont{V.}~\bibnamefont{Malka}}, \bibinfo {author}
  {\bibfnamefont{A.}~\bibnamefont{Tafzi}}, \bibinfo {author}
  {\bibfnamefont{J.~P.}\ \bibnamefont{Goddet}}, \bibinfo {author}
  {\bibfnamefont{R.~C.}\ \bibnamefont{Shah}}, \bibinfo {author}
  {\bibfnamefont{S.}~\bibnamefont{Sebban}},\ and\ \bibinfo {author}
  {\bibfnamefont{A.}~\bibnamefont{Rousse}},\ }%
  \bibfield{journal}{%
  \Doi{10.1038/nphoton.2012.82}{\bibinfo {journal} {Nat. Photon.}}\ }%
  \textbf{\bibinfo {volume} {6}},\ \bibinfo {pages} {308Ð311} (\bibinfo {month}
  {March}\ \bibinfo {year} {2012})%
  \bibAnnoteFile{NoStop}{MalkaNature}%
\bibitem{RPastro}%
  \BibitemOpen
  \bibfield{author}{%
  \bibinfo {author} {\bibfnamefont{G.~B.}\ \bibnamefont{{Rybicki}}}\ and\
  \bibinfo {author} {\bibfnamefont{A.~P.}\ \bibnamefont{{Lightman}}},\ }%
  \emph{\bibinfo {title} {Radiative Processes in Astrophysics}}\ (\bibinfo
  {publisher} {John Wiley and Sons},\ \bibinfo {year} {1979})%
  \bibAnnoteFile{NoStop}{RPastro}%
\bibitem{LWFAscailing}%
  \BibitemOpen
  \bibfield{author}{%
  \bibinfo {author} {\bibfnamefont{W.}~\bibnamefont{Lu}}, \bibinfo {author}
  {\bibfnamefont{M.}~\bibnamefont{Tzoufras}}, \bibinfo {author}
  {\bibfnamefont{C.}~\bibnamefont{Joshi}}, \bibinfo {author}
  {\bibfnamefont{F.~S.}\ \bibnamefont{Tsung}}, \bibinfo {author}
  {\bibfnamefont{W.~B.}\ \bibnamefont{Mori}}, \bibinfo {author}
  {\bibfnamefont{J.}~\bibnamefont{Vieira}}, \bibinfo {author}
  {\bibfnamefont{R.~A.}\ \bibnamefont{Fonseca}},\ and\ \bibinfo {author}
  {\bibfnamefont{L.~O.}\ \bibnamefont{Silva}},\ }%
  \bibfield{journal}{%
  \Doi{10.1103/PhysRevSTAB.10.061301}{\bibinfo {journal} {Phys. Rev. ST Accel.
  Beams}}\ }%
  \textbf{\bibinfo {volume} {10}},\ \bibinfo {pages} {061301} (\bibinfo {month}
  {Jun}\ \bibinfo {year} {2007})%
  \bibAnnoteFile{NoStop}{LWFAscailing}%
\bibitem{NikishovRitus}%
  \BibitemOpen
  \bibfield{author}{%
  \bibinfo {author} {\bibfnamefont{A.~I.}\ \bibnamefont{Nikishov}}\ and\
  \bibinfo {author} {\bibfnamefont{V.~I.}\ \bibnamefont{Ritus}},\ }%
  \bibfield{journal}{%
  \bibinfo {journal} {Sov. Phys. JETP}\ }%
  \textbf{\bibinfo {volume} {19}},\ \bibinfo {pages} {529} (\bibinfo {year}
  {1964})%
  \bibAnnoteFile{NoStop}{NikishovRitus}%
\bibitem{kogaPOP}%
  \BibitemOpen
  \bibfield{author}{%
  \bibinfo {author} {\bibfnamefont{J.}~\bibnamefont{Koga}}, \bibinfo {author}
  {\bibfnamefont{T.~Z.}\ \bibnamefont{Esirkepov}},\ and\ \bibinfo {author}
  {\bibfnamefont{S.~V.}\ \bibnamefont{Bulanov}},\ }%
  \bibfield{journal}{%
  \Doi{10.1063/1.2013067}{\bibinfo {journal} {Physics of Plasmas}}\ }%
  \textbf{\bibinfo {volume} {12}},\ \bibinfo {eid} {093106} (\bibinfo {year}
  {2005})%
  \bibAnnoteFile{NoStop}{kogaPOP}%
\bibitem{OSIRIS}%
  \BibitemOpen
  \bibfield{author}{%
  \bibinfo {author} {\bibfnamefont{R.~A.}\ \bibnamefont{Fonseca}}, \bibinfo
  {author} {\bibfnamefont{L.~O.}\ \bibnamefont{Silva}}, \bibinfo {author}
  {\bibfnamefont{F.~S.}\ \bibnamefont{Tsung}}, \bibinfo {author}
  {\bibfnamefont{V.~K.}\ \bibnamefont{Decyk}}, \bibinfo {author}
  {\bibfnamefont{W.}~\bibnamefont{Lu}}, \bibinfo {author}
  {\bibfnamefont{C.}~\bibnamefont{Ren}}, \bibinfo {author}
  {\bibfnamefont{W.~B.}\ \bibnamefont{Mori}}, \bibinfo {author}
  {\bibfnamefont{S.}~\bibnamefont{Deng}}, \bibinfo {author}
  {\bibfnamefont{S.}~\bibnamefont{Lee}}, \bibinfo {author}
  {\bibfnamefont{T.}~\bibnamefont{Katsouleas}},\ and\ \bibinfo {author}
  {\bibfnamefont{J.~C.}\ \bibnamefont{Adam}},\ }%
  \emph{\bibinfo {title} {Lect Notes Comput Sc}},\ Vol.\ \bibinfo {volume}
  {2331}\ (\bibinfo {publisher} {Springer Berlin / Heidelberg},\ \bibinfo
  {year} {2002})\ pp.\ \bibinfo {pages} {342--351}%
  \bibAnnoteFile{NoStop}{OSIRIS}%
\bibitem{myAPS}%
  \BibitemOpen
  \bibfield{author}{%
  \bibinfo {author} {\bibfnamefont{M.}~\bibnamefont{Vranic}}, \bibinfo {author}
  {\bibfnamefont{J.~L.}\ \bibnamefont{Martins}},\ and\ \bibinfo {author}
  {\bibfnamefont{L.~O.}\ \bibnamefont{Silva}},\ }%
  \bibfield{journal}{%
  \bibinfo {journal} {Bulletin of the American Physical Society}\ }%
  \textbf{\bibinfo {volume} {54}},\ \bibinfo {pages} {CP8 71} (\bibinfo {month}
  {Nov}\ \bibinfo {year} {2009}),\
  \url{http://meetings.aps.org/Meeting/DPP09/Event/108980}%
  \bibAnnoteFile{NoStop}{myAPS}%
\bibitem{Samuel_nature}%
  \BibitemOpen
  \bibfield{author}{%
  \bibinfo {author} {\bibfnamefont{S.~F.}\ \bibnamefont{Martins}}, \bibinfo
  {author} {\bibfnamefont{R.~A.}\ \bibnamefont{Fonseca}}, \bibinfo {author}
  {\bibfnamefont{W.}~\bibnamefont{Lu}}, \bibinfo {author}
  {\bibfnamefont{W.~B.}\ \bibnamefont{Mori}},\ and\ \bibinfo {author}
  {\bibfnamefont{L.~O.}\ \bibnamefont{Silva}},\ }%
  \bibfield{journal}{%
  \Doi{10.1038/nphys1538}{\bibinfo {journal} {Nature Phys.}}\ }%
  \textbf{\bibinfo {volume} {6}},\ \bibinfo {pages} {311} (\bibinfo {month}
  {March}\ \bibinfo {year} {2010})%
  \bibAnnoteFile{NoStop}{Samuel_nature}%
\bibitem{Schwinger}%
  \BibitemOpen
  \bibfield{author}{%
  \bibinfo {author} {\bibfnamefont{J.}~\bibnamefont{Schwinger}},\ }%
  \bibfield{journal}{%
  \Doi{10.1103/PhysRev.82.664}{\bibinfo {journal} {Phys. Rev.}}\ }%
  \textbf{\bibinfo {volume} {82}},\ \bibinfo {pages} {664} (\bibinfo {month}
  {Jun}\ \bibinfo {year} {1951})%
  \bibAnnoteFile{NoStop}{Schwinger}%
\bibitem{Joana}%
  \BibitemOpen
  \bibfield{author}{%
  \bibinfo {author} {\bibfnamefont{J.~L.}\ \bibnamefont{Martins}}, \bibinfo
  {author} {\bibfnamefont{S.~F.}\ \bibnamefont{Martins}}, \bibinfo {author}
  {\bibfnamefont{R.~A.}\ \bibnamefont{Fonseca}},\ and\ \bibinfo {author}
  {\bibfnamefont{L.~O.}\ \bibnamefont{Silva}},\ }%
  \bibfield{journal}{%
  \Doi{10.1117/12.820736}{\bibinfo {journal} {Harnessing Relativistic Plasma
  Waves as Novel Radiation Sources from Terahertz to X-Rays and Beyond}}\ }%
  \textbf{\bibinfo {volume} {7359}},\ \bibinfo {pages} {73590V} (\bibinfo
  {year} {2009})%
  \bibAnnoteFile{NoStop}{Joana}%
\end{thebibliography}%

\end{document}